\documentclass[prl,twocolumn,aps,showpacs,superscriptaddress]{revtex4}
\usepackage{graphics}
\usepackage{epsfig}
\usepackage{amsmath}
\usepackage{amssymb}
\usepackage{bm}
\usepackage{color}
\begin{document}

\title{Shear dilatancy in marginal solids
}
\author{Brian P. Tighe
}

\affiliation{Delft University of Technology, Process \& Energy Laboratory, Leeghwaterstraat 44, 2628 CA Delft, The Netherlands}

\date{\today}

\begin{abstract}
Shearing stresses can change the volume of a material via a nonlinear effect known as shear dilatancy. We calculate the elastic dilatancy coefficient of  soft sphere packings and random spring networks, two canonical models of  marginal solids close to their rigidity transition. We predict a dramatic enhancement of  dilatancy near rigidity loss in both materials, with a surprising distinction: while packings expand under shear, networks contract. We show that contraction in networks is due to the destabilizing influence of increasing hydrostatic or uniaxial loads, which is counteracted in packings by the formation of new contacts.
\end{abstract}

\maketitle

More than a century ago, Reynolds noted the tendency of granular materials to increase their volume in response to shear stresses \cite{reynolds1885}. Today the phenomenon still fascinates \cite{onoda90,pouliquen96,luding05,schroeter07,daniels08,gravish10,metayer11}. Recent work by Behringer and co-workers \cite{ren13} has sparked interest in the confluence of dilatancy and (un)jamming, the nonequilibrium rigidity transition in disordered soft matter \cite{ohern03,vanhecke10}. Their experiments probed  a corollary to Reynolds dilatancy in which normal stresses are induced by shear at constant volume.
They found that the normal stresses induced in sheared granular materials increase dramatically on approach to the critical volume fraction where hydrostatically compressed packings jam. Similar phenomena are seen in quasistatically sheared foams \cite{weaire03}.

Yet there is no rigorous mechanical bound requiring sheared isotropic solids to expand, and in fact positive shear dilatancy is not a universal feature of elastic response. Stiff biopolymer networks sheared at constant volume, for example, develop  an ``anti-granular'' negative normal stress  -- they pull rather than push on shearing surfaces  \cite{janmey07,conti09}. 

\begin{figure}[tb]
\centering 
\includegraphics[clip,width=0.85\linewidth]{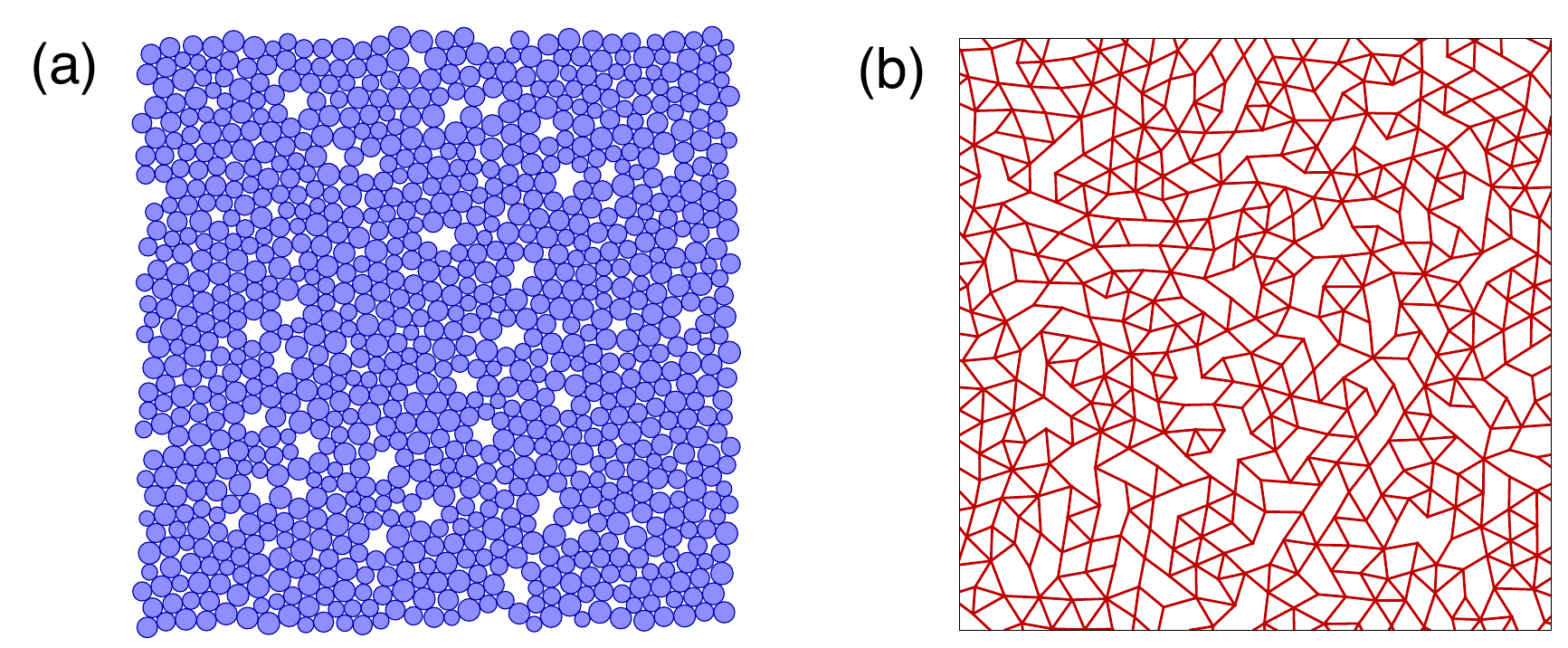}
\caption{Two model marginal solids: (a) a frictionless soft sphere packing; (b) a random spring network.}
\label{fig:systems}
\end{figure}

What determines if a sheared material initially expands or contracts? 
Here we show that dilatancy in marginal elastic solids is closely tied to the evolution of its shear modulus under compressive loading.
We present a calculation of the  dilatancy coefficient of two closely related models of marginal solids (Fig.~\ref{fig:systems}), both of which undergo an unjamming transition at a critical network connectivity and zero load. They are (i) random networks of linear springs, whose connectivity is independent of their load; and (ii) packings of soft frictionless spheres jammed by a confining pressure $p_0 > 0$, whose connectivity is an increasing function of the pressure. We find that random networks  soften under increased pressure and therefore contract under shear. Sphere packings stiffen as compression creates new contacts; as a result, they expand when sheared. 

\begin{figure}[tb]
\centering 
\includegraphics[clip,width=0.75\linewidth]{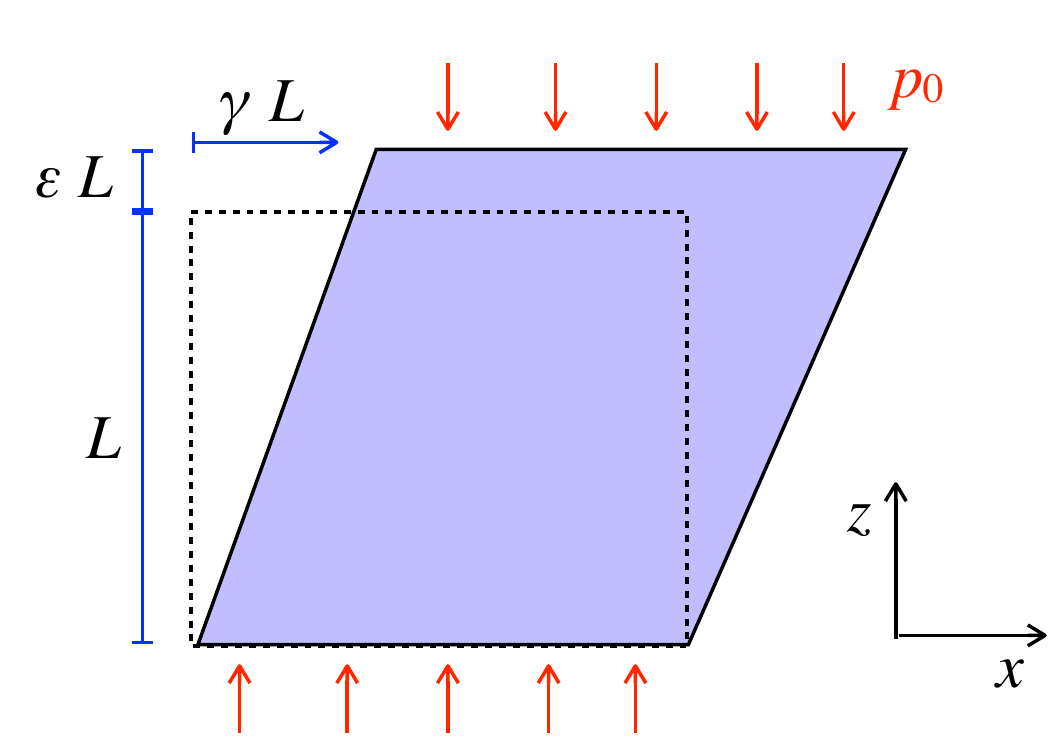}
\caption{Shear dilatancy. A material (dashed box) of linear extent $L$ initially under pressure $p_0$ (arrows; horizontal normal stresses omitted) is subjected to a shear strain $\gamma$ while the normal stress on the shearing surface is held constant. A dilatant strain $\epsilon \propto \gamma^2$ results.}
\end{figure}

\section{Elastic shear dilatancy}
We begin by relating the dilation of a sheared elastic solid to its linear response to shear and compression.

We treat quasistatic elastic deformations of the parallelepiped in Fig.~1 \cite{landau}. For simplicity, we consider an initial state that is a cube with cross-sectional area $A = L^2$, initial height $h_0 = L$, and initial volume $V_0 = A h_0$. We further consider only deformations holding $A $ fixed. Points $\vec r$ on the boundary displace by a distance
$\delta {\vec r} = \gamma  r_x \, {\hat {\rm e}_x} + \epsilon  r_z \,   {\hat {\rm e}_z}$,
where ${\hat {\rm e}_x}$ and  ${\hat {\rm e}_z}$ are unit vectors and
the parameters $\gamma$ and $\epsilon$ are shear and dilatant strains, respectively. The stress conjugate to $\gamma$ is the shear stress $\sigma$, which is zero in the initial condition. The stress conjugate to $\epsilon$ is the normal stress $p$; its value in the initial state is the pressure $p_0$.

We first consider a system in which the normal stress is held constant during shearing. The case of constant volume ($\epsilon = 0$) is treated below. 

The dilatant strain $\epsilon$ can be Taylor expanded in powers of he shear strain $\gamma$,
\begin{equation}
\epsilon  = D_p \gamma + \frac{1}{2}R_p \gamma^2 + O(\gamma^4) \,,
\label{eqn:quadratic}
\end{equation}
The linear coefficient $D_p$ is zero in isotropic materials, which must dilate by the same amount whether sheared to the left or the right.\footnote{In disordered and statistically isotropic materials of finite size, $D_p$ will vanish with increasing system size.}
The coefficient of the quadratic term in Eq.~(\ref{eqn:quadratic}) is the Reynolds coefficient,
\begin{equation}
R_p = \lim_{\gamma \rightarrow 0} \, \left( \frac{\partial^2 \epsilon}{\partial \gamma^2} \right)_p \,.
\end{equation}

\paragraph{Calculating the Reynolds coefficient.} The Reynolds coefficient $R_p$ can be related to a material's linear elastic moduli via a simple derivation first reported by Weaire and Hutzler in the context of liquid foams \cite{weaire03}. Here we follow a different route to compatible conclusions.

Quasistatic linear response of an isotropic elastic solid is characterized by two moduli; it will prove convenient to select the shear modulus $G$ and the Young's modulus $E$. The shear modulus quantifies the proportionality between shear stress and shear strain $\gamma$ in the small strain limit,
\begin{equation}
G = - \lim_{\gamma \rightarrow 0} \left( \frac{\partial \sigma}{\partial \gamma} \right)_\epsilon  \,.
\label{eqn:G}
\end{equation}
The Young's modulus quantifies the change in normal stress that results from dilatant strain,
\begin{equation}
E = - \lim_{\epsilon \rightarrow 0} \left( \frac{\partial p}{\partial \epsilon} \right)_\gamma  \,.
\label{eqn:E}
\end{equation}

Our starting point assumes a hyperelastic solid and an initially isotropic state with finite $G$ and $E$. Its differential elastic potential energy is ${\rm d}U = -p A \, {\rm d}h - \sigma A h \, {\rm d}\gamma$, with $h = (1 + \epsilon)h_0$.
To treat stress-controlled boundary conditions it is useful to introduce the Legendre transform $H = U + pAh$, which is a natural function of $p$ and $\gamma$:
\begin{equation}
{\rm d}H = Ah\, {\rm d}p - \sigma A h \, {\rm d}\gamma \,.
\label{eqn:enthalpy}
\end{equation}
The  Maxwell relation corresponding to ${\rm d}H$ is
\begin{equation}
\left( \frac{\partial h}{\partial \gamma} \right )_p = - \left( \frac{\partial (\sigma h)}{\partial p} \right )_\gamma  \,.
\label{eqn:maxwell}
\end{equation}
Invoking Eqs.~(\ref{eqn:quadratic}), (\ref{eqn:G}), and (\ref{eqn:E}) and equating the leading order terms of Eq.~(\ref{eqn:maxwell}) gives an explicit relation between the dilatancy coefficient and the two linear elastic moduli,
\begin{equation}
R_p =   \left( \frac{\partial G}{\partial p} \right)_\gamma - \frac{G}{E}   \,.
\label{eqn:Rp}
\end{equation}

Let us consider the consequences of Eq.~(\ref{eqn:Rp}). Mechanical stability imposes positivity on both $G$ and $E$, but the variation of the shear modulus with  load can be either positive or negative. Hence the balance of terms in Eq.~(\ref{eqn:Rp}) can have either sign -- an elastic material can either expand or contract under shear. 

In the two model marginal solids treated below, 
\begin{equation}
\frac{G}{E} < 1 
\label{eqn:GEratio}
\end{equation}
and
\begin{equation}
\left | \left(\frac {\partial G}{\partial p} \right)_\gamma \right | \gg 1 \,,
\label{eqn:susceptible}
\end{equation}
hence
\begin{equation}
R_p \simeq   \left(\frac {\partial G}{\partial p} \right)_\gamma \,
\label{eqn:Rscaling}
\end{equation}
close to unjamming. Eq.~(\ref{eqn:GEratio})  is quite general, as nearly all naturally occurring materials are stiffer against compression than shear.  Moreover, strong susceptibility to external forcing is commonly observed near a phase transition, so we expect Eqs.~(\ref{eqn:susceptible}) and (\ref{eqn:Rscaling}) to hold in  other marginal solids. 

To summarize, a marginal solid sheared at constant normal stress expands if its shear modulus is a strongly increasing function of $p$. It contracts if its shear modulus is a strongly decreasing function of $p$. This is the first of our three main results.

\begin{figure}[tb]
\centering 
\includegraphics[clip,width=1.0\linewidth]{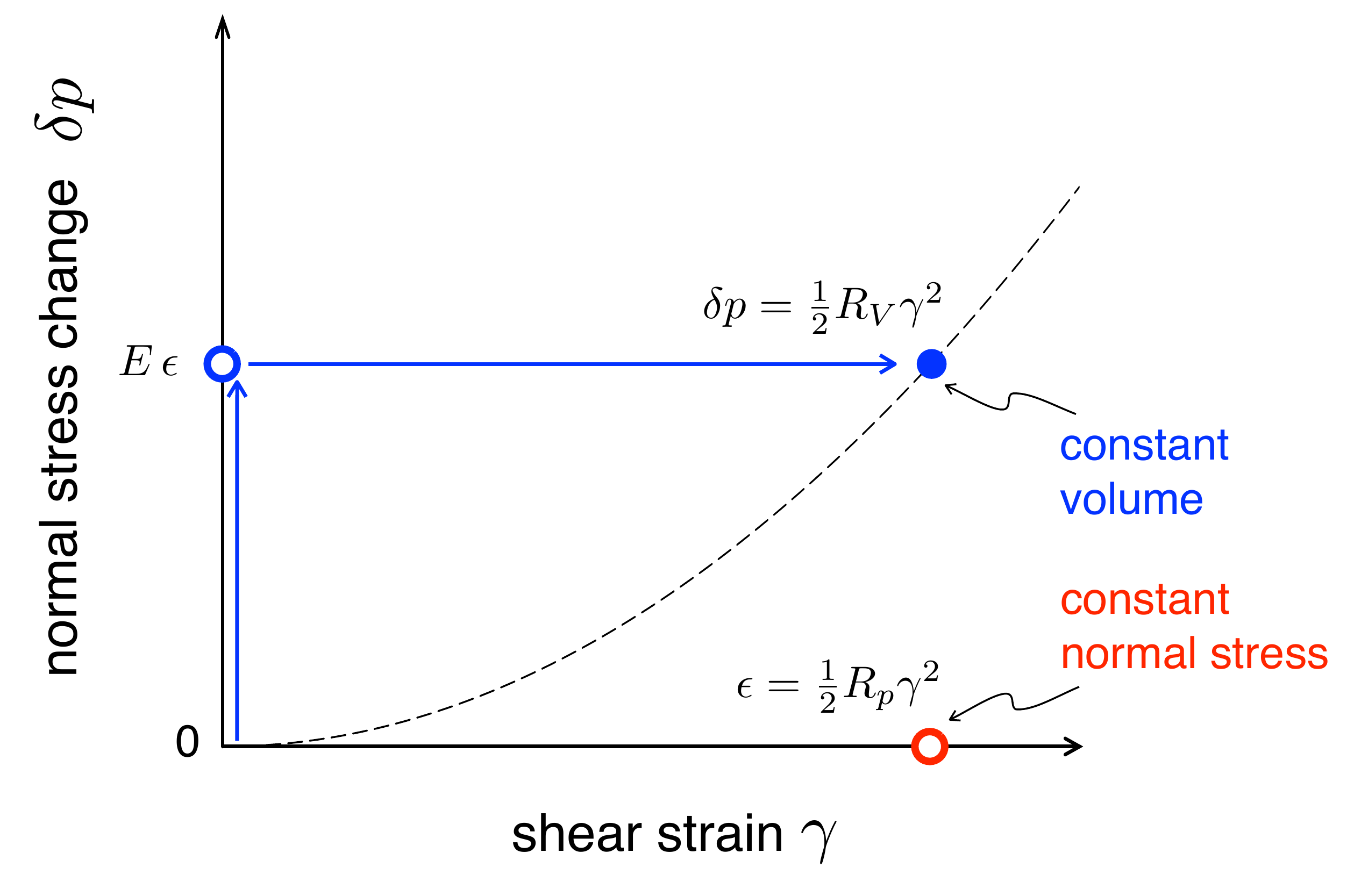}
\caption{Relating shear at constant volume and constant normal stress.}
\end{figure}

\paragraph{Shearing at fixed volume.} Before analyzing the Reynolds coefficient in two model materials, we consider the case of shear at constant volume. When the volume is held fixed instead of the normal stress, the same symmetry arguments invoked above require a normal stress change $\delta p = p - p_0$ having the form
\begin{equation}
\delta p  = \frac{1}{2}R_V \gamma^2 + O(\gamma^4) \,.
\end{equation}
The coefficient $R_V$ is
\begin{equation}
R_V = \lim_{\gamma \rightarrow 0} \,  \left( \frac{\partial^2 p}{\partial \gamma^2} \right)_\epsilon \,.
\end{equation}

The  coefficient $R_V$ can be evaluated in the same way as $R_p$. Alternatively, one may view the system's isochoric trajectory in Fig.~2 as the sum of two processes: (i) uniaxial expansion, and (ii)  shear to strain $\gamma$ at constant normal stress. The dilatant strain $\epsilon = -[(1/2)R_p \gamma^2]$ imposed in the first process is chosen to counterbalance that induced by the second process.\footnote{Shearing at $p_0 + \delta p$ instead of $p_0$ introduces a negligible correction of order $\gamma^4$.} Thus
\begin{equation}
R_V = E R_p = -V_0 \left( \frac{\partial G}{\partial V} \right )_{\gamma, A} - G\,.
\label{eqn:RV}
\end{equation}
As in the stress-controlled case, the first term will dominate in marginal solids.

An important and intuitive consequence of Eq.~(\ref{eqn:RV}) is that $R_p$ and $R_V$ have the same sign, so one can speak unambiguously of `positive' and `negative' shear dilatancy without reference to the boundary conditions. A material that expands (contracts) when sheared at constant normal stress will increase (decrease) its normal stress when sheared at constant volume.

\section{Dilatancy in networks and packings}
 \begin{figure}[tb]
\centering 
\includegraphics[clip,width=1\linewidth]{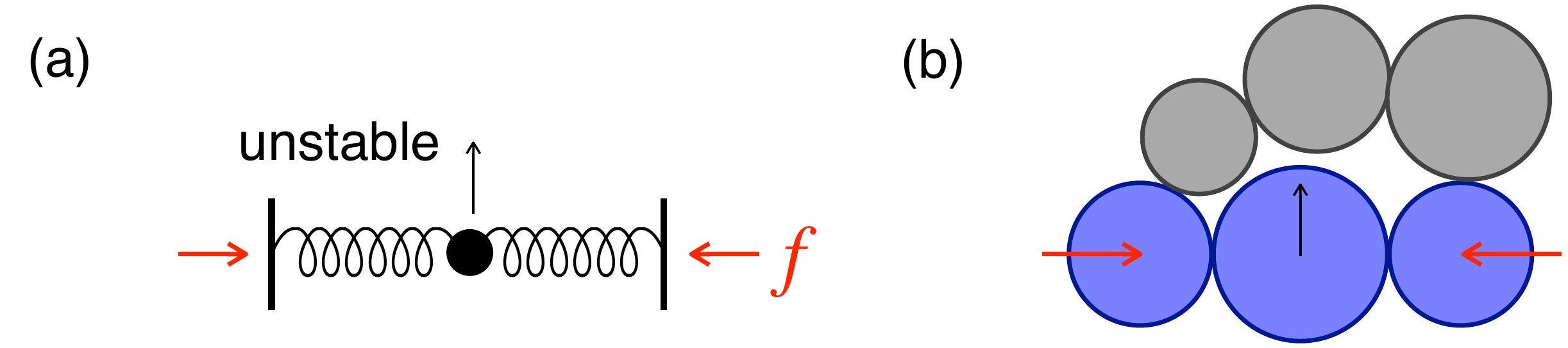}
\caption{(a) A segment from a loaded chain of springs. The chain collapses under compressive loading. (b) The analog of the loaded chain in a soft disk packing. Collapse is stabilized by the formation of new contacts}
\label{fig:buckling}
\end{figure}

We now consider shear dilatancy in two model marginal solids: random spring networks \cite{wyart08,ellenbroek09b,tighe12c,tighe08a} 
and soft frictionless sphere packings \cite{ohern03,tighe11} (Fig.~\ref{fig:systems}).

 Our analysis will hinge on the observation that spring networks are destabilized by increasing normal load, while packings are stabilized.
 In a general sense, this distinction is intuitive: while compression induces buckling in elastic structures \cite{landau}, packings owe their rigidity to a positive confining pressure \cite{dagois-bohy12,alexander}.
Further illustration can be provided by the loaded chain, a familiar example from mechanics. Fig.~\ref{fig:buckling}a depicts one segment of a chain. It is straightforward to show that a chain is stable to transverse motions when the longitudinal load $f$ is below a threshold $f^*$. For this simple example the threshold is at $f^* = 0$, i.e.~the chain collapses under any compressive load. Fig.~\ref{fig:buckling}b depicts the soft sphere analog of Fig.~\ref{fig:buckling}a embedded in a packing. In displacing transversely, the central disk forms a new contact. This process continues until the packing achieves a topology and geometry capable of supporting the load. Hence contact formation enhances stability to increased loading. The competing effects of softening due to compressive loading and stiffening due to contact formation were first emphasized by Alexander \cite{alexander}.

\paragraph{Negative dilatancy in networks.} 
We first treat random networks of harmonic springs with a mean connectivity of $z$ springs per node. Absent loading, such networks are rigid above a critical connectivity $z_c$ which in mean field is given by Maxwell's isostatic value $2d$, where $d$ is the dimension. 

\begin{figure}[tb]
\centering 
\includegraphics[clip,width=0.9\linewidth]{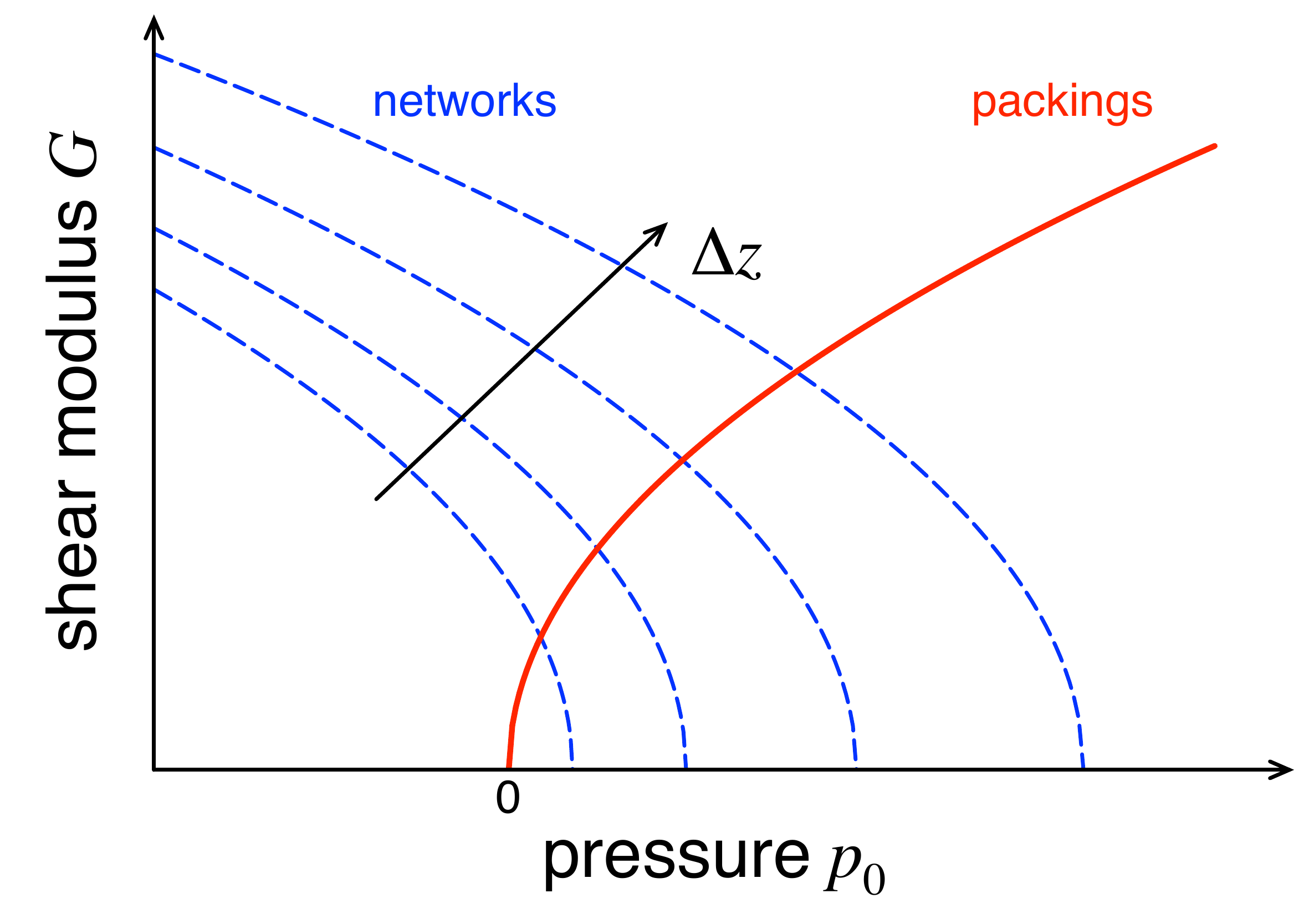}
\caption{Evolution of the shear modulus $G$ of random networks (dashed curves, varying proximity to isostaticity $\Delta z$) and soft sphere packings (solid curve) as a function of the hydrostatic load $p_0$.  Increased loading softens the shear response of networks but stiffens that of packings. The slope of $G$ controls dilatancy, hence networks contract while packings expand. }
\label{fig:theory}
\end{figure}

The shear modulus of a spring network can be calculated from a normal mode analysis, the details of which are omitted here but can be found in Refs.~\cite{tighe11,tighe12b}. The result is that $G$ is directly proportional to the frequency $\omega^*$ that characterizes a plateau of low frequency ``soft modes'' that appears in the vibrational density of states close to isostaticity \cite{wyart05b}.  One finds
\begin{equation}
G \sim \Delta z \sqrt{1- c \, \frac{p_0}{\Delta z^2}} \,.
\label{eqn:Gnet}
\end{equation}
Here $\Delta z = z -z_c \ge 0$ is the distance to isostaticity and $c$ is a positive constant.  The shear modulus as a function of load is depicted in Fig.~\ref{fig:theory} for several different connectivities. The key observation is that $G$ {\rm softens} with increasing load, ultimately vanishing at a load $p^* \sim \Delta z^2$ where the network becomes unstable to shear. While some network preparation protocols display critical exponents that deviate from the mean field prediction of Eq.~(\ref{eqn:Gnet}) \cite{broedersz11}, we expect  softening with increasing loading to be generic.

Invoking Eq.~(\ref{eqn:Rscaling}), the scaling of the dilatancy coefficient is 
\begin{equation}
R_p \sim -\frac{1}{\Delta z} \left( 1- c \, \frac{p_0}{\Delta z^2} \right)^{-1/2} < 0\,,
\label{eqn:networkR}
\end{equation}
which reduces to 
\begin{equation}
R_p \sim -\frac{1}{\Delta z} \,,
\label{eqn:networkR2}
\end{equation}
for an initially unloaded network.
Here and below we assume that the distinction between hydrostatic and uniaxial loading can be neglected for the purposes of scaling analysis.

The shear and Young's moduli of unloaded networks have a constant ratio, $G/E \sim {\rm const}$ \cite{ellenbroek09b}, hence their Reynold's coefficient at fixed volume is 
\begin{equation}
R_V \sim {\rm const} < 0 \,.
\label{eqn:networkRV}
\end{equation}

Eqs.~(\ref{eqn:networkR} - \ref{eqn:networkRV}) are our second main result: dilatancy in marginal spring networks is negative, so they {\em contract} under shear. Physically, this is a consequence of the destabilizing influence of increased normal load. 
For constant normal stress, moreover, $R_p$ diverges on approach to the critical connectivity at zero load.

\paragraph{Positive dilatancy in soft spheres.} 

As a counterpoint to random spring networks we now consider packings of frictionless spheres (Fig. \ref{fig:systems}a), which undergo an unjamming transition when the confining pressure is sent to zero. Overlapping spheres repel elastically with a contact potential $U \propto \delta^\alpha$, where $\delta$ is the dimensionless overlap. Here we treat the harmonic case $\alpha = 2$, which is an appropriate model for foams and emulsions. The generalization to the Hertzian case for solid spheres, $\alpha = 5/2$,  is straightforward.

The shear modulus of harmonic soft sphere packings is known to scale as the square root of pressure \cite{ohern03},
\begin{equation}
G \sim p_0^{1/2} \,.
\label{eqn:Gpacking}
\end{equation}
This scaling relation is illustrated in Fig.~\ref{fig:theory}. 
The scaling of the dilatancy coefficient follows immediately from (\ref{eqn:Rscaling}),
\begin{equation}
R_p \sim \frac{1}{p_0^{1/2}} > 0\,.
\label{eqn:Rpacking}
\end{equation}

Unlike networks, the Young's modulus of a packing approaches a finite constant near the jamming transition \cite{ohern03}. As a result, their Reynolds coefficient at constant volume is also diverging,
\begin{equation}
R_V \sim \frac{1}{(\phi - \phi_c)^{1/2}} \,.
\label{eqn:RVpacking}
\end{equation}
Here $\phi$ is the volume fraction of the packing and $\phi_c$ is the critical volume fraction where packings unjam.

Eqs.~(\ref{eqn:Rpacking}) and (\ref{eqn:RVpacking}) are our final main result. They show that shear dilatancy in marginally solid packings is diverging and positive -- sheared packings {\em expand}. 

Physically, the softening experienced by the spring networks is avoided in packings because, unlike networks, packings can create new contacts. While pressure and connectivity can be varied independently in networks, the (excess) connectivity of a sphere packing is selected by the pressure, $\Delta z = \Delta z(p_0)$. 
The shear modulus of networks can  be used to rationalize the  form of $\Delta z(p_0)$. To see this, we begin from the shear modulus of Eq.~(\ref{eqn:Gnet}) and demand a mechanically stable state, which must satisfy $G>0$. The result is a bound on the connectivity of a sphere packing:
\begin{equation}
\Delta z > \sqrt{c \, p_0} \,.
\label{eqn:inequality}
\end{equation}
This inequality, first derived by Wyart et al.~\cite{wyart05b}, indicates that packings at higher pressures require more contacts to remain stable.
Marginal solids saturate the bound: simulations, foams, and grains all display square root scaling of their excess connectivity  \cite{ohern03,katgert10b,majmudar07},
\begin{equation}
\Delta z \sim p_0^{1/2} \,.
\label{eqn:zscaling}
\end{equation}
Inserting Eq.~(\ref{eqn:zscaling}) in Eq.~(\ref{eqn:Gnet}), one recovers the shear modulus of soft sphere packings, Eq.~(\ref{eqn:Gpacking}).

Note that the shear modulus of any single, finite-sized packing will display discontinuities as new contacts are formed under compression. The smooth square root scaling of $G$, which dictates $R_p$, holds in the limit of thermodynamically large systems or under ensemble averaging.

\section{Discussion}

Shear dilatancy results from a subtle, nonlinear coupling between the responses to compression and shear. We have shown positive  dilatancy occurs in marginal solids that stiffen against shear when compressed, while negative dilatancy is found in those materials that soften. 

Our calculations provide a unifying perspective for the seemingly disparate dilatant phenomena found in frictionless materials such as emulsions and liquid foams (which expand) and semiflexible biopolymer networks (which contract). Compression stiffens the shear response of soft sphere packings by creating new contacts, and hence packings expand under shear. Because random spring networks cannot create new contacts, they soften under compression and contract under shear.

The scaling relations of Eqs.~(\ref{eqn:networkR}) and (\ref{eqn:Rpacking}) require numerical confirmation. Tests in sheared networks and  packings are underway and will be reported elsewhere.

Our calculations for soft sphere packings describe the approach to unjamming in frictionless materials. The frictionless jamming point occurs at a critical packing fraction $\phi_c$ commonly known as random close packing (RCP); we describe the approach to RCP from above ($\phi > \phi_c$).
Unlike frictionless sphere packings, frictional states can exist at zero pressure over a range of packing fractions of which RCP is the upper bound. The Behringer experiments approach RCP from below, beginning from isotropic states at (nearly) zero pressure \cite{ren13}. Any relation between these two approaches to RCP is not  {\em a priori} apparent. Nor is it self-evident that 
 frictional packings can be modeled as hyperelastic solids.
These issues can be explored numerically in frictional soft sphere packings \cite{somfai07}.

\begin{acknowledgements}
I acknowledge valuable discussions with Bob Behringer and Joshua Dijksman.
\end{acknowledgements}

\bibliographystyle{spphys}       

\begin{thebibliography}{10}
\providecommand{\url}[1]{{#1}}
\providecommand{\urlprefix}{URL }
\expandafter\ifx\csname urlstyle\endcsname\relax
  \providecommand{\doi}[1]{DOI \discretionary{}{}{}#1}\else
  \providecommand{\doi}{DOI \discretionary{}{}{}\begingroup
  \urlstyle{rm}\Url}\fi

\bibitem{reynolds1885}
O.~Reynolds, in \emph{Proc. Brit. Assoc.} (1885), p. 896

\bibitem{onoda90}
G.Y. Onoda, E.G. Liniger, Phys. Rev. Lett. \textbf{64}, 2727 (1990)

\bibitem{pouliquen96}
O.~Pouliquen, N.~Renaut, Journal de Physique II \textbf{6}(6), 923 (1996)

\bibitem{luding05}
S.~Luding, J. Phys. Cond. Matt. \textbf{17}, S2623 (2005)

\bibitem{schroeter07}
M.~Schr\"oter, S.~N\"agle, C.~Radin, H.L. Swinney, EPL \textbf{78}(4), 44004
  (2007)

\bibitem{daniels08}
K.E. Daniels, N.W. Hayman, J. Geophys. Res. \textbf{113}, B11411 (2008)

\bibitem{gravish10}
N.~Gravish, P.B. Umbanhowar, D.I. Goldman, Phys. Rev. Lett. \textbf{105},
  128301 (2010)

\bibitem{metayer11}
J.F. M\'etayer, D.J. Suntrup, C.~Radin, H.L. Swinney, M.~Schr\"oter, EPL
  \textbf{93}, 64003 (2011)

\bibitem{ren13}
J.~Ren, J.A. Dijksman, R.P. Behringer, Phys. Rev. Lett. \textbf{110}, 018302
  (2013)

\bibitem{ohern03}
C.S. O'Hern, L.E. Silbert, A.J. Liu, S.R. Nagel, Phys.~Rev.~E \textbf{68},
  011306 (2003)

\bibitem{vanhecke10}
M.~van Hecke, J.~Phys.~Cond.~Matt. \textbf{22}, 033101 (2010)

\bibitem{weaire03}
D.~Weaire, S.~Hutzler, Philosoph.~Mag. \textbf{83}, 2747 (2003)

\bibitem{janmey07}
P.A. Janmey, M.E. McCormick, S.~Rammensee, J.L. Leight, P.C. Georges, F.C.
  MacKintosh, Nature Materials \textbf{6}, 48  (2007)

\bibitem{conti09}
E.~Conti, F.C. MacKintosh, Phys. Rev. Lett. \textbf{102}, 088102 (2009)

\bibitem{landau}
L.D. Landau, E.M. Lifshitz, \emph{Theory of Elasticity} (Butterworth-Heineman,
  Oxford, 1997)

\bibitem{wyart08}
M.~Wyart, H.~Liang, A.~Kabla, L.~Mahadevan, Phys. Rev. Lett. \textbf{101},
  215501 (2008)

\bibitem{ellenbroek09b}
W.G. Ellenbroek, Z.~Zeravcic, W.~van Saarloos, M.~van Hecke, Europhys. Lett.
  \textbf{87}, 34004 (2009)

\bibitem{tighe12c}
B.P. Tighe, Phys. Rev. Lett. \textbf{109}, 168303 (2012)

\bibitem{tighe08a}
B.P. Tighe, J.E.S. Socolar, Phys.~Rev.~E \textbf{77}, 031303 (2008)

\bibitem{tighe11}
B.P. Tighe, Phys. Rev. Lett. \textbf{107}, 158303 (2011)

\bibitem{dagois-bohy12}
S.~Dagois-Bohy, B.P. Tighe, J.~Simon, S.~Henkes, M.~van Hecke, Phys. Rev. Lett.
  \textbf{109}, 095703 (2012)

\bibitem{alexander}
S.~Alexander, Phys.~Rep \textbf{296}, 65 (1998)

\bibitem{tighe12b}
B.P. Tighe, {\tt arXiv:1205.2960v1}  (2012)

\bibitem{wyart05b}
M.~Wyart, L.E. Silbert, S.R. Nagel, T.A. Witten, Phys.~Rev.~E \textbf{72},
  051306 (2005)

\bibitem{broedersz11}
C.P. Broedersz, X.~Mao, T.C. Lubensky, F.C. MacKintosh, Nat. Phys. \textbf{7},
  983 (2011)

\bibitem{katgert10b}
G.~Katgert, M.~van Hecke, EPL \textbf{92}(3), 34002 (2010)

\bibitem{majmudar07}
T.S. Majmudar, M.~Sperl, S.~Luding, R.P. Behringer, Phys.~Rev.~Lett.
  \textbf{98}, 058001 (2007)

\bibitem{somfai07}
E.~Somfai, M.~van Hecke, W.G. Ellenbroek, K.~Shundyak, W.~van Saarloos,
  Phys.~Rev.~E \textbf{75}, 060302(R) (2007)

\end{thebibliography}

\end{document}